# Nuclear spin population and its control toward initialization using an all-electrical sub-micron scale nuclear magnetic resonance device

T. Ota
*NTT Basic Research Laboratories, NTT Corporation, 3-1, Morinosato-Wakamiya, Atsugi, 243-0198, Japan and SORST-JST, 4-1-8 Honmachi, Kawaguchi, Saitama 331-0012, Japan*

G. Yusa
*NTT Basic Research Laboratories, NTT Corporation, 3-1, Morinosato-Wakamiya, Atsugi, 243-0198, Japan; PRESTO-JST, 4-1-8 Honmachi, Kawaguchi, Saitama 331-0012, Japan; and Department of Physics, Tohoku University, Sendai, 980-8578, Japan*

N. Kumada
*NTT Basic Research Laboratories, NTT Corporation, 3-1, Morinosato-Wakamiya, Atsugi, 243-0198, Japan*

S. Miyashita
*NTT-AT, 3-1, Morinosato-Wakamiya, Atsugi, 243-0198, Japan*

Y. Hirayama
*NTT Basic Research Laboratories, NTT Corporation, 3-1, Morinosato-Wakamiya, Atsugi, 243-0198, Japan; SORST-JST, 4-1-8 Honmachi, Kawaguchi, Saitama 331-0012, Japan; and Department of Physics, Tohoku University, Sendai, 980-8578, Japan*

We study the nuclear spin population in a GaAs quantum well structure and demonstrate its initialization using an all-electrical nuclear magnetic resonance (NMR) device. In our device, nuclear spins are dynamically polarized in a sub-micron scale region defined by split gates. The nuclear spin populations under various polarization conditions are estimated from resistively-detected pulsed NMR spectra. We find that nuclear spin populations are determined by electron spin configurations. By applying radio frequency pulses to the strongly polarized nuclear spins, we demonstrate the creation of two-qubit effective pure states, which is a crucial step toward NMR quantum computation.

Nuclear magnetic resonance (NMR) techniques that allow the implementation of quantum computation and information processing have attracted considerable interest recently. One of the crucial steps for performing quantum algorithms is the initialization of nuclear spins, i.e. the creation of effective pure states, which transform in a similar way to pure states under unitary operations [1]. In standard NMR, the initialization is accomplished by applying radio frequency (RF) pulses to thermally populated nuclear spins. However, this process is inefficient because the number of effective nuclei contributing to such states is quite small. Recently, it has been demonstrated that highly populated nuclear spins can be obtained by current-driven dynamic polarization using an integer/fractional quantum Hall device [2,3]. These devices enable the manipulation of nuclear spins by all-electrical means, and have considerable potential advantages such as scalability and capacity for combination with conventional electronics.

We focus on such a fractional quantum Hall device in which nuclear spins are dynamically polarized in a sub-micron scale region defined by split gates and manipulated by RF pulses applied through an antenna gate. In this device, the nuclear spin states after manipulation can be sensitively detected as the change in the longitudinal resistance $\Delta R_{xx}$ which relates to the change in the $z$ component of magnetization $\Delta M_z$ of the nuclear spins [3]. By sweeping the frequency of the RF pulse while measuring $\Delta R_{xx}$, resistively-detected NMR spectra are obtained. In this Letter, we first examine the population of four nuclear spin levels of $^{75}$As and $^{69}$Ga each with nuclear spin $I=3/2$ under various polarization conditions, and show that the population depends heavily on the electronic spin configuration. Then, by selecting a starting condition with appropriately polarized nuclear spins, we demonstrate the initialization into two-qubit effective pure states.

Figure 1(a) schematically shows the cross-sectional view of the device used in this study. We fabricated a Hall bar using a material consisting of a 20 nm-thick GaAs/Al$_{0.3}$Ga$_{0.7}$As quantum well. A back gate enables the control of the electron density ($N_S$) of the two-dimensional electron gas (2DEG), which is formed in the GaAs layer. The mobility is typically 300 m$^2$/Vs for $N_S=1.0\times10^{15}$ m$^{-2}$. A single pair of metal split gates with a gap of 800 nm is deposited on the surface in order to squeeze the conducting 2DEG in the GaAs layer into a narrow channel. By applying a negative voltage (-0.4 V) and a constant source-drain current (10-15 nA), we can locally increase the current density between the split gates. The electron-nuclear spin coupling is pronounced only in this constricted region when an external static magnetic field $B_0$ is applied to set the electron system at Landau level filling factor $\nu=2/3$ [3]. At this filling factor, two electron spin states, i.e. spin-polarized and unpolarized states, are formed in the 2DEG. The dynamic nuclear spin polarization occurs due to flip-flop scattering upon transport across boundaries of spin-polarized and unpolarized domains, giving rise to the hyperfine field which affects the 2DEG as an additional source of the scattering [4]. Then, the polarization is subsequently detected as an enhancement of $R_{xx}$. For the manipulation of nuclear spins, RF pulses are irradiated through an antenna gate which is deposited just above the split gates [3]. The manipulated nuclear spin states are sensitively detected as $\Delta R_{xx}$. In our device, $\Delta R_{xx}$ directly relates to the change in the $z$ component of magnetization $\Delta M_z$. This is in contrast to the standard NMR in which a pick-up coil detects an oscillating voltage induced by the precession of the transverse magnetization $M_{xy}$ [5]. The $R_{xx}$ measurements are performed using standard lock-in techniques. The sample is mounted in a dilution refrigerator with a temperature below 150 mK.

Figure 1(b) shows a resistively-detected pulsed NMR spectrum for $^{75}$As nuclei at 4 T, taken by sweeping the frequency of the RF pulse while measuring $\Delta R_{xx}$. The detail measurement sequence of the resistively-detected pulsed NMR spectrum is described in Ref. 2. The three peaks correspond to possible single-photon transitions between four spin levels $|3/2\rangle, |1/2\rangle, |-1/2\rangle$ and $|-3/2\rangle$ of $^{75}$As nuclei with nuclear spin $I=3/2$. The peak splitting is determined by the electric quadrupolar constant $E_Q$ since the transition energy can be expressed as $\gamma\hbar B_0 - 6E_Q$, $\gamma\hbar B_0$ and $\gamma\hbar B_0 + 6E_Q$ [5]. Here, $\hbar$ is Planck's constant $h$ divided by $2\pi$, and $\gamma$ is the gyromagnetic ratio. $E_Q$ can be either positive or negative depending on the electric field distortion, which breaks the symmetry at the nuclear sites [5]. Therefore, the relative energy of the four spin levels depends on the sign of $E_Q$. In this Letter, we consider the case of $E_Q>0$ for both $^{69}$Ga and $^{75}$As nuclei since the essence of the discussion is not altered by the sign of $E_Q$. For $E_Q>0$, the peaks labeled *I*, *II* and *III* correspond to transitions *I*, *II* and *III*, respectively, in the inset of Fig. 1(b).

In order to analyze the relative population difference of the four spin levels after dynamic polarization, we focus on the intensity of the NMR spectrum. In this study, the peak intensity is detected after applying a 130 µsec-long pulse, which approximately corresponds to a $\pi$ pulse for all transitions *I*, *II* and *III* (we call this pulse a detection pulse). As shown in Fig. 1(b), the peak intensity is highly asymmetric, indicating a low effective nuclear spin temperature [5]. We employ simple numerical analysis to estimate the occupation ratios of each nuclear spin state [6]. In this calculation, using $\Delta R_{xx}=\alpha M_z$, we estimate the change in $M_z$ after the irradiation of the detection pulse. Here, $\alpha$ is a conversion coefficient, and its unit is $\Omega$ per nuclear magneton $\mu_N$. The occupation ratios can be defined as $\rho = \rho^0 + \rho^1$ where $\rho = \rho^0 = 0.25$ under equilibrium before dynamic polarization and $\rho^1$ is the deviation induced by dynamic polarization. For the spectrum in Fig. 1(b), the numerical analysis yields $(\rho^1_{3/2}, \rho^1_{1/2}, \rho^1_{-1/2}, \rho^1_{-3/2}) = (-882/4\alpha, -667/4\alpha, 54/4\alpha, 1494/4\alpha)$. The relation $\rho^1_{3/2} < \rho^1_{1/2} < \rho^1_{-1/2} < \rho^1_{-3/2}$ implies a negative spin temperature. The dashed line in Fig. 1(b) is the calculated spectrum using the rotating-frame approximation method [3], in which $\rho_{3/2}(B_0=4T)$ is taken to be zero. This condition in the calculation leads to $\alpha = 882$ $\Omega/\mu_N$ and the numerical values for $(\rho_{3/2}, \rho_{1/2}, \rho_{-1/2}, \rho_{-3/2}) = (0, 0.06, 0.27, 0.67)$ after dynamic polarization, indicating a spin temperature of $-1$ mK. The calculated spectrum using these values well reproduces the experimental one as shown in Fig. 1(b).

We now examine what effect changing $B_0$ has on the nuclear spin population. $B_0$ can be changed without altering $\nu$ by suitably adjusting the back gate voltage to maintain $\nu = N_S/(eB_0/h) = 2/3$. Here, $e$ is the elementary charge. Figure 2(a) shows NMR spectra for $^{75}$As taken at $B_0=4$ T to 8 T. The frequencies of the peaks in the NMR spectra are stated with respect to peak *II*. The integrated total intensity of the spectrum is maximal at 4 T, indicating the strongest polarization. This intensity gradually decreases as $B_0$ increases as shown in Fig. 2(b). The relative amplitudes of the peaks change significantly, i.e. the intensity of the right-hand peak is maximal in the spectra at $B_0=4$ T to 6 T, whereas that of the left-hand peak is strongest at $B_0=7$ T to 8 T. The change in the relative peak intensity results from a change of $\rho^1$ during dynamic polarization. By fitting the experimental spectra with the calculated ones, $\rho^1$ can be extracted. Figure 2 (c) shows $\rho^1$ at $B_0=4$ T to 8 T, normalized by $|\rho^1_{3/2}(B_0=4T)|$. The occupation probability $\rho^1_{-3/2}$ is 1.69 at 4 T, and gradually decreases to 0.4 as $B_0$ increases up to 8 T [7]. On the other hand, $\rho^1_{1/2}$ increases from -0.76 at 4 T to -0.02 at 8 T. $\rho^1$ is determined by the competition between the polarization and depolarization rates of nuclear spins during dynamic polarization through the mechanisms such as flip-flop scattering, relaxation, diffusion and so on. [8]. We expect that the electron spin configuration plays an important role for the determination of these rates. It is because, at $\nu=2/3$, it is known that there are two electron spin configurations, i.e. spin-polarized and unpolarized states which are favoured in the $B_0$ region higher and lower than 6 T, respectively [4]. Another possible candidate for the determination of the rates is $B_0$ which yields the level spacing $\gamma\hbar B_0$. However, the effect of $B_0$ can be discarded from the fact that a similar change in the relative peak intensity as shown in Fig. 2(a) is observed for a fixed $B_0 =6$ T, where spin-polarized and unpolarized states coexist and the corresponding spectra can be taken by only changing $N_S$ using the back gate. From these results, we can conclude that a change in the electron spin configuration leads to different conditions for dynamic polarization, resulting in a significant change in $\rho^1$. We speculate that the depolarization rate is mainly modified by the electron spin configurations since they strongly affect the relaxation of nuclear spins [9].

We now demonstrate the creation of two-qubit effective pure states using strongly polarized nuclear spins [10]. Here, the four spin levels $|3/2\rangle, |1/2\rangle, |-1/2\rangle$ and $|-3/2\rangle$ of As (or Ga) can be considered as $|00\rangle, |01\rangle, |10\rangle$ and $|11\rangle$ states of the two-qubit system [11]. In this experiment, we start from the condition where the population linearly increases with spin states, resulting in a spectrum with nearly equivalent peak heights as shown in Fig. 3(a). The solid and dashed lines in Fig. 3(a) correspond to the experimental spectrum and the calculated one for $^{69}$Ga at 6 T, respectively. The 180 µsec-long detection pulse was applied to take this spectrum. The filled circles in the schematics indicate spin populations in the four levels after dynamic polarization. Starting from the population corresponding to the spectrum in Fig. 3(a), two consecutive pulses are used to create an effective pure state $|11\rangle$ ($|00\rangle$). The first one is a $\pi$ pulse on transition *I* (*II*), which exchanges the population between states $|00\rangle$ and $|01\rangle$ ($|01\rangle$ and $|10\rangle$), and the second is a $\pi/2$ pulse on transition *II* (*III*) that equalizes the population in states $|01\rangle$ and $|10\rangle$ ($|11\rangle$ and $|10\rangle$). The spectrum in Fig. 3(b) (3(c)) corresponds to the effective pure state $|11\rangle$ ($|00\rangle$) measured by applying these two consecutive pulses followed by the detection pulse. The effective pure state $|10\rangle$ ($|01\rangle$) can be created by applying an additional $\pi$ pulse to transition *III* (*I*) after the pulse sequence to create the effective pure state $|11\rangle$ ($|00\rangle$), resulting in the spectrum in Fig. 3(d) (3(e)). In this experiment, all the pulses are applied with no time delay. The dashed lines in Figs. 3(b) to 3(e) are calculated spectra assuming that the populations estimated from the spectrum in Fig. 3(a) are properly transferred by the pulse sequences, and well-reproduce the experimental data. The obtained spectra for the effective pure states $|10\rangle$ and $|01\rangle$ in Figs. 3(d) and 3(e) slightly deviate from the calculated spectra around 60.635 and 60.665 MHz, respectively. This deviation is caused by additional population transfer during the irradiation of the detection pulse. The pulse sequences described here are valid for only the linearly populated nuclear spin states. However, by employing a more sophisticated pulse sequence, we can make similar effective pure states starting from any arbitrary nuclear spin populations of $^{75}$As and $^{69,71}$Ga nuclei at various $B_0$.

In our scheme based on $M_z$ detection, only diagonal elements in the density matrix can be detected [12]. Therefore, it is necessary to consider off-diagonal components in the density matrix separately after the pulse sequences. In this study, we use the π/2 pulse which creates unwanted transverse magnetization. In standard NMR, before reading the final results, a pulsed field gradient is usually applied in order to destroy the transverse magnetization, i.e. average out all of the off-diagonal terms in the density matrix [10]. On the other hand, in our device, we set a time interval of 5 sec between the detection pulse and the $R_{xx}$ measurement. This interval is significantly long compared with the dephasing time $T_2$ of 0.6 msec but within the spin-lattice relaxation time $T_1$ of a few tens of minutes. Then, at the time of the measurement, all off-diagonal terms are negligible due to decoherence. Therefore, the effect of the off-diagonal components in the density matrix can be safely discarded, and this confirms initialization of nuclear spins as shown in Figs. 3(b)-(e).

In case of demonstrating quantum algorithms, the density matrix has to be reconstructed by quantum-state tomography techniques [1]. In order to measure the off-diagonal terms using our device based on $M_z$ detection, they have to be transferred to the diagonal components within $T_2$, and then read out them within $T_1$ by applying the pulses [12].

In conclusion, we have studied the nuclear spin populations of constituent nuclei in a GaAs quantum well under different dynamic polarization conditions at $\nu = 2/3$. We estimated the occupation ratios of nuclear spins for the four spin levels. We found that the electron spin configuration plays an important role for determining the nuclear spin population. Using the linearly populated nuclear spin states, we created effective pure states by applying RF pulses.

The authors thank K. Muraki, A. Miranowicz, K. Takashina, S. K. Ozdemir, N. Imoto, T. Fujisawa, and P. Balestriere for valuable discussions.

[Figure captions]

FIG.1 (a) Schematic illustration of a cross-sectional view of the device. (b) Resistively-detected NMR spectrum of $^{75}$As nuclei at 4 T. The duration of detection pulse is 130 μsec. The quadrupolar splitting between the center peak and satellites is 28 kHz. The dashed line is the calculated spectrum. Inset: Schematic energy level diagram of nuclear spins for *I*=3/2 with the electric quadrupolar constant $E_Q$>0. The transitions *I*, *II* and *III* in the inset correspond to the peaks labeled as *I*, *II* and *III*.

FIG. 2. (a) Pulsed NMR spectra for $^{75}$As at various $B_0$ and back gate voltages. The frequencies of the peaks in the NMR spectra are stated with respect to peak *II*. The duration of detection pulse is 130 μsec. The spectra at $B_0$ =4 to 8 T are taken at back gate voltages of -0.59, -0.37, -0.14, 0.05 and 0.27 V, respectively. (b) Integrated intensity of the spectra at $B_0$ =4 T to 8 T. (c) $\rho^1$ at $B_0$ =4 T to 8 T, normalized by $|\rho^1_{3/2}(B_0 = 4T)|$.

FIG. 3. (a) Pulsed NMR spectrum of $^{69}$Ga nuclei at 6 T. The quadrupolar splitting is 14 kHz. The duration of detection pulse is 180 μsec. The dotted line is the calculated spectrum. (b)-(e) The obtained spectra corresponding to effective pure states $|11\rangle, |00\rangle, |10\rangle, |01\rangle$, respectively. The dotted lines are calculated spectra assuming that the populations are properly transferred by the PF pulses. The filled circles show the schematic spin populations in the four levels.

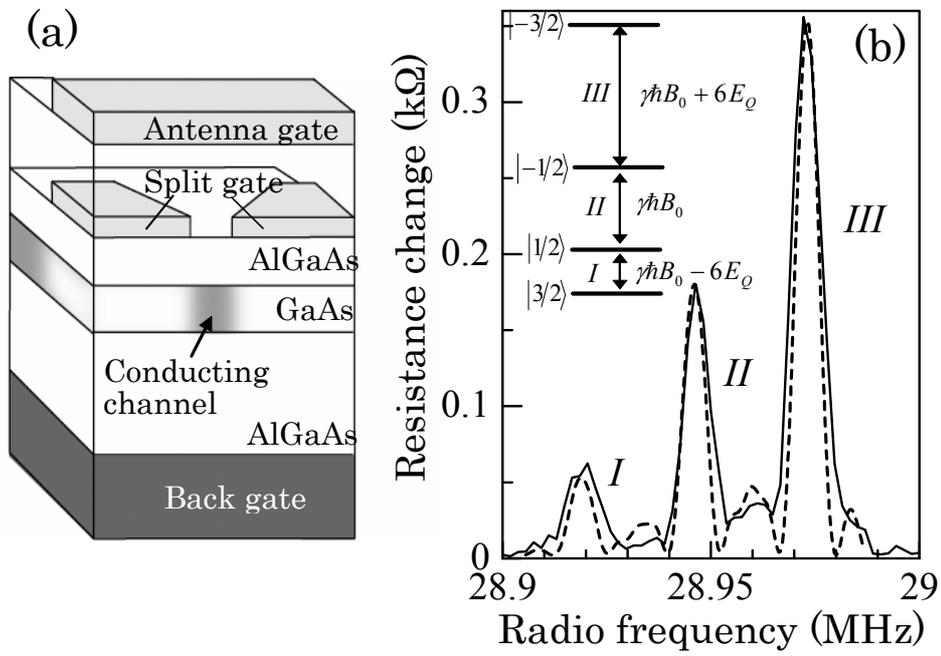

Fig. 1

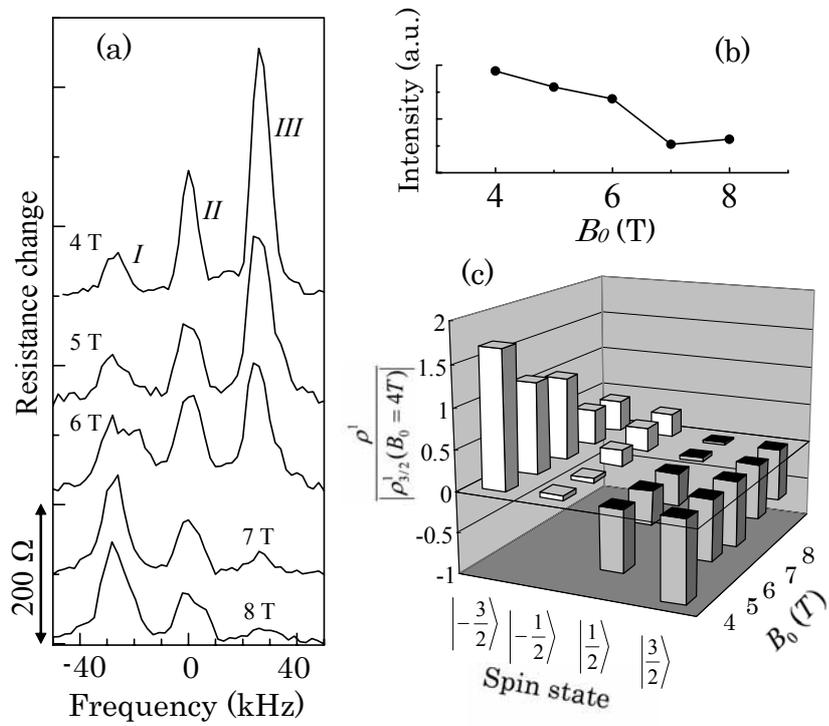

Fig. 2

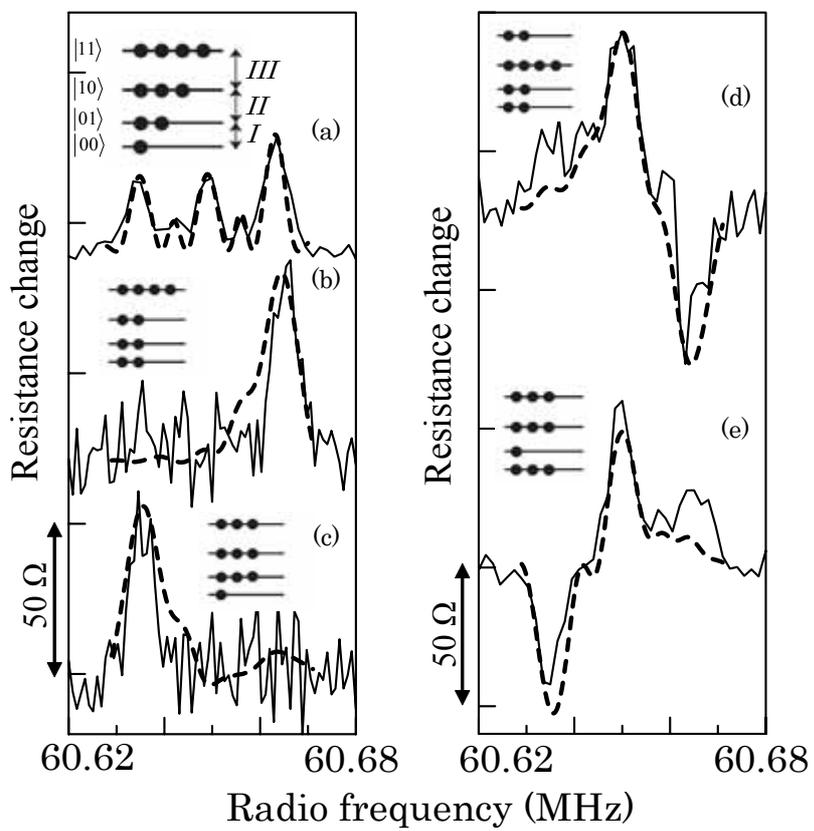

Fig. 3